\documentclass{ws-ijmpd}
\usepackage[super,compress]{cite}
\begin{document}


%
\catchline{}{}{}{}{}
%

\title{Primordial black hole collision with neutron stars and astrophysical black 
holes and the observational signatures}

\author{Sohrab Rahvar}

\address{Physics Department, Sharif University of Technology, Tehran 11365-9161,  Iran \\
and \\
 Research Center for High Energy Physics, Sharif University of Technology, Tehran, Iran.
rahvar@sharif.edu}

\maketitle

\begin{history}
\received{Day Month Year}
\revised{Day Month Year}
\end{history}

\title

\maketitle

\begin{history}
\received{Day Month Year}
\revised{Day Month Year}
\end{history}

\begin{abstract}
In this paper, we examine whether low-mass Primordial Black Holes (PBHs) can be considered a plausible dark matter candidate in galactic halos. We derive the relativistic dynamics of PBHs around the heavy compact objects and evaluate their collision rate, as well as the likelihood of PBH capture in neutron stars and black holes. Although the rate of these collisions in the Milky Way is lower than our lifetime (i.e. almost one collision per hundred years), it may still be observable on cosmological scales. Additionally, we investigate the gravitational wave emission as an important observable window for PBH-astrophysical black hole merging. For the allowed range of PBH mass, gravitational wave signal is smaller than the sensitivity of present gravitational wave detectors. We provide observational prospect for detection of these events in future. 

\end{abstract}


\section{\label{introduction} Introduction}

The large-scale structures of the universe are dominantly composed of dark matter and Primordial black holes (PBHs) are one of the potential candidates for dark matter, believed to have formed in the early universe due to quantum fluctuations \cite{zeldovich,hawking}. While the theory of PBH formation predicts a wide range of masses for PBHs \cite{kuhnel}, observations constrain their mass to two narrow windows: low mass ($M<10^{-7} M_\odot$) and high mass ($M>1000 M_\odot$) \cite{green,moniez2}. It is believed that galaxies are surrounded by a halo of dark matter, which we assume to be comprised of PBHs.


In this work, our aim is to investigate the interaction of PBHs with astrophysical compact objects. This study has been done in numerous works such as PBH merging with the astrophysical structures through the adiabatic process \cite{derishev,Fabio,loeb,2019JCAP,2020PhRvD,Esser}. Also, the collision of PBH with stars and compact objects are modeled as GRB source \cite{Graham}. The consequences of this collision as the Gamma-ray emission were also  examined with the X-ray observatories \cite{abram}.  

Here, we focus on investigating the rate of PBH collisions with neutron stars and astrophysical black holes (hereafter referred to as compact objects) and studying the consequences of this interaction. While the PBH collision with Earth has been studied in the context of Newtonian gravity \cite{rahvar}, we extend this argument to PBH collisions with compact objects with relativistic dynamics, considering astrophysical consequences of collision of the PBHs with the neutron stars and the astrophysical blackholes and possible observational signatures from this collisions.



In Section (\ref{s1}), we introduce the phase space of PBHs and calculate the relativistic path of PBHs around the compact objects and the rate of collisions that occur with compact objects. In Section (\ref{sec3}), we study the consequences of these collisions with neutron stars, and in Section (\ref{sec4}), we investigate the collision of PBHs with astrophysical black holes. Finally, Section (\ref{sec5}) provides a summary and conclusion of our study.


\section{Collision of PBHs with compact objects}
\label{s1}
In this section, we study the phase space of PBHs in the Galactic halo and the rate of collisions with compact objects.
\subsection{Dynamics of PBHs in halo}
Let us assume the distribution of point mass PBHs in the halo of the Milky Way galaxy is almost isothermal and follows the Maxwell-Boltzmann distribution as 
\begin{equation}
\label{max}
f(x,v) d^3v = n_0(x) (\frac{3}{2\pi\sigma^2})^{3/2}\exp{(\frac{-3v^2}{2\sigma^2})}v^2dv d\Omega,
\end{equation}
where $"x"$ represents the spatial part of phase space and $n_0(x)$ is the number density of PBHs at any point of space.  $d\Omega = d\phi\sin\theta d\theta$ is the solid angle element and $\sigma$ is the dispersion velocity of PBHs. Let us assume a coordinate system where a compact object is set at the center of it.

This object gravitationally interacts with the PBHs in the halo. The flux of particles in this coordinate system (at far distance of "$r$" from the center) entering the sphere with the radial distance of $"r"$ is \cite{spergel} 
\begin{equation}
\label{current}
{\cal F}(r) = \int  f(r,v) {\bf\hat n}\cdot {\bf v}  v^2 dv d\Omega, 
\end{equation}
where ${\hat n}$ is the inwarding unit vector perpendicular to the surface and $0<\theta<\pi/2$.
We note that the overall current in the steady state condition for the incoming and outgoing particles is zero.  Here, our aim is to consider just the incoming PBHs inside the sphere with the radius of $r$.   Multiplying the flux in equation (\ref{current}) to the area at far distances from the center, results in the rate of particles crossing the sphere with the radius of $r$ (i.e. $dN/dt = 4\pi r^2 {\cal F}$). 

We define the kinetic energy and the angular momentum of PBHs at asymptotically flat spacetime with respect to the center of coordinate as $E = mv^2/2$ and $J = rmv\sin\theta$.  We note that the compact object at the center of this coordinate system has a non-zero velocity with an arbitrary direction with respect to the halo. For an ensemble of compact objects in the Galaxy we can ignore the peculiar velocity of compact object while a detailed calculation would provide a dipole interaction-rate of PBHs with the compact object at the center of coordinate. 

We can calculate the rate of incoming particles inside a sphere with a radius of $r$ in terms of energy and angular momentum as 
\begin{equation}
\label{rate0}
\frac{dN}{dt} = 2\pi^2 n_0 (\frac{3}{2\pi m^2\sigma^2})^{3/2}\int \exp(-\frac{3E}{m\sigma^2})dEdJ^2.
\end{equation}

In the next section we derive the trajectory of PBHs around the gravitational potential of the central compact object in terms of the two conserved quantities of  $E$ and $J$. Then our aim is to obtain the rate of interactions of PBHs with the central compact object as a function of these two conserved quantities.


\subsection{Trajectory of PBHs  around compact objects: Relativistic approach}
Let us assume the Schwarzschild metric for the spacetime around a compact object as
\begin{equation}
    ds^2 = - (1-\frac{r_s}{r})c^2dt^2 + (1-\frac{r_s}{r})^{-1}dr^2 + r^2 d\Omega.
\end{equation}
 The Schwarzschild radius is $r_s = 2GM/c^2$. 


The action for a test particle  of mass $m$ \cite{landau} (here a point mass PBH around the compact object) is 
\begin{equation}
  S = -mc^2\int d\tau = \int {\cal L} dt 
  \end{equation}
  where 
  \begin{equation}
    {\cal L} = - mc^2\sqrt {(1-\frac{r_s}{r}) - (1-\frac{r_s}{r})^{-1}\dot{r}^2 - r^2\sin^2\theta\dot{\phi}^2-r^2\dot{\theta}^2}.
\end{equation}
For simplicity in the calculation, we can reduce space dimension into two, taking into account the initial condition of the particle  by setting $\dot{\theta} = 0$ and $\theta = \pi/2$. The corresponding canonical momentum for each coordinate is given by $p_i = \partial {\cal L}/\partial \dot{q}_i$ where from the Lagrangian the momentum is  
\begin{eqnarray}
    p_r &=& -\frac{m^2c^2\dot{r}}{{\cal L}}(1-\frac{r_s}{r})^{-1}, \\
    J &=& -\frac{m^2c^2r^2\dot{\phi}}{\cal{L} },
\end{eqnarray}
where $J$ is a conserved quantity. 
The corresponding Hamiltonian is 
${\cal H} = \sum p_i \dot{q}_i - \cal{L}$ where substituting the  momentum and coordinate, the Hamiltonian simplifies to 
\begin{equation}
    {\cal H} = - \frac{m^2c^4}{\cal{L}}(1-\frac{r_s}{r}).
\end{equation}
Dividing $J$ by the Hamiltonian we can find the angular velocity in terms of these two quantities,  
\begin{equation}
    \dot{\phi} = \frac{J c^2}{\cal H}\frac{(r-r_s)}{r^3}.
\end{equation}
Since Hamiltonian is time-independent, that is a conserved quantity and we define it at the asymptotic flat space-time 
\begin{eqnarray}
\label{h}
        {\cal H} &=& \frac{mc^2}{\sqrt{1-v^2}}\simeq mc^2 + E_k,\\  
   &\lim&~~r\rightarrow \infty \nonumber 
\end{eqnarray}
where $E_k = \frac12 m v_0^2$ is the kinetic energy of PBH in the Galactic halo.

On the other hand, let us take the closest distance of the test particle to a compact object that happens at $r=r_{min}$ where at this position the radial velocity is zero. Then the Hamiltonian at this point is given by 
\begin{equation}
\label{h2}
    {\cal H}^2 =(1-\frac{r_s}{r_{min}})m^2c^4 + 
    \frac{J^2c^2}{r_{min}^2}.
\end{equation}
Equating the left and right-hand sides of equations (\ref{h}) and (\ref{h2}), then the angular momentum in terms of the kinetic energy of PBH at far distance and $r_{min}$ is 
\begin{equation}
\label{J}
    J^2 =m^2 c^2 r_{min}^2\left((1+\frac{E_k}{mc^2})^2 + \frac{r_s}{r_{min}} - 1\right),
\end{equation}
where for the Newtonian limit (i.e. $r_s\rightarrow 0 $) we recover the conventional angular momentum as $J = r_{min} m v(r_{min})$. 
This equation represents a constraint between the angular momentum and kinetic energy of a particle.
If the radius of a compact object, denoted as $r_c$ is greater than the minimum distance between the compact object and a PBH, denoted as $r_{min}$ (i.e. $r_c>r_{min}$), then the collision between the two objects occur at $r_c$. 

\subsection{The rate of collisions}
For calculating the rate of collisions, we substitute equation (\ref{J}) as the constrain between the energy and angular momentum of PBHs in (\ref{rate0}), and after integration the result is 
\begin{equation}
\label{rate}
\frac{dN}{dt} =2\pi^2 n_0(\frac{3}{2\pi})^{3/2}\sigma r_{min}^2\left(\frac{2}{9} + \frac{2}{3}\frac{r_s} {r_{min}}\frac{c^2}{\sigma^2} + \frac{2}{27}\frac{\sigma}{c}\right),
\end{equation}
where the first and second terms are Newtonian terms; the third term is the relativistic correction to the collision rate. Taking the local density of the dark halo around the disk of the Milky Way \cite{binney}  as $\rho_D \simeq 8\times 10^{-3} M_\odot pc^{-3}$  and the mass range of PBHs as $[10^{14},10^{23}]$gr, the number density of PBHs in the halo obtain 
 $$   n_0 = 6\times 10^{-33} f \times(\frac{\rho_h}{0.008 M_\odot/pc^{3}})(\frac{m_{pbh}}{10^{23} g})^{-1}  \text{km}^{-3}$$
where $f$ is the fraction of dark matter made of PBHs. Then the numerical value of equation (\ref{rate}) is 
\begin{eqnarray}
\label{rate2}
   \frac{dN}{dt} &=& 2.7 \times 10^{-11}\text{Gyr}^{-1} f\times(\frac{\rho_h}{0.008 M_\odot/pc^{3}})(\frac{m_{pbh}}{10^{23} g})^{-1} \\ 
     & & \times\left(0.22
    + 2.25\times 10^5(\frac{rs}{km})(\frac{r_m}{10 km})^{-1}\sigma_{200}^{-2} + 5\times 10^{-5}\sigma_{200} \right),\nonumber
\end{eqnarray}
 where $\sigma_{200}$ is the dispersion velocity of PBHs in the halo normalized to $200$km/s. Ignoring the relativistic terms, the result agrees with the Newtonian calculation \cite{abram}. For the collision of PBHs with compact objects in the halo of Galaxy, we can ignore the first and third terms compared to the second term; however for the dense regions as the center of Galaxy where the PBHs move with relativistic velocities, we have to keep all the terms. For the non-relativistic velocities the rate simplifies to 
    \begin{equation}
    \label{rate4}
     \frac{dN}{dt} = 6.1\times 10^{-6} \text{Gyr}^{-1} f\times (\frac{m_{pbh}}{10^{23} g})^{-1} (\frac{rs}{km})(\frac{r_m}{10 km})^{-1}\sigma_{200}^{-2}.
    \end{equation}

Taking into account the evaporation of PBHs from the early universe up to the present time, we expect PBHs with the masses of $m>10^{14}$gr could survive at the present time \cite{rahvar}. For the PBHs with Dirac-Delta mass function within the mass range of $m\in[10^{14},10^{23}]$g \cite{carr,green} (which is marginally larger than the experimental constrain) the rate of collisions from equation (\ref{rate4}) is obtained as $dN/dt\in[6~ ,6\times 10^{-9}]\text{Myr}^{-1}$. While the collision rate of PBH with a compact object is high, however, the rate of capturing (by means of trapping inside the compact object or making a gravitationally bound state) is low. 
In the next sections, we investigate the capture rate and the astrophysical signatures from the PBH collisions with compact objects.

\section{Physics of PBH collision with Neutron stars}
\label{sec3} In this section, 
 we investigate the physical consequences of the PBH collision with compact objects such as neutron stars. Let us image a small mass PBH colliding with a neutron star where we can ignore the  momentum transfer to the neutron star as the mass of neutron star is much larger than the mass of PBHs. Once PBH enters the neutron star two physical processes can happen (i) decelerating due to the dynamical friction. (ii) accretion of mass of neutron star on PBHs. For a PBH colliding with a neutron star, PBH interacts gravitationally with the condensed material inside the neutron star. The result is a drag force so-called dynamical friction \cite{chand} which is given by 
\begin{equation}
\frac{d\mathbf{v}_{pbh}}{dt} = -\frac{4\pi \ln (\Lambda) G^2  \rho_n m_{pbh}}{v_{pbh}^3}\left[\mathrm{erf}(X)-\frac{2X}{\sqrt{\pi}}e^{-X^2}\right]\mathbf{v}_{pbh},
\label{dfric}
\end{equation}
where $\rho_n$ is the average density of a neutron star, $m_{pbh}$ is the mass of PBH as a projectile, $v_{pbh}$ is the velocity of PBH inside the neutron star that is derived from equation (\ref{h2}), $ X = v_{pbh}/(\sqrt{2} \sigma_n)$ and $\sigma_n$ is the dispersion velocity of particles inside the neutron star. $\Lambda$ is given by 
 \begin{equation}
 \Lambda = \frac{r_n \sigma_n^2}{Gm_{pbh}} = 2(\frac{r_n}{r_{s(bh)}})(\frac{\sigma_n}{c})^2, 
 \label{lam}
 \end{equation}
 where $r_{n}$ is the radius of the neutron star, $r_{s(pbh)}$ is the Schwartzchild radius of PBH. 
 Substituting the numerical values:
 \begin{equation}
     \Lambda = 6 \times 10^{10} (\frac{r_n}{10 \text{km}})(\frac{m_{pbh}}{10^{23}\text{g}})^{-1}(\frac{\sigma_n}{c^2})^2,
 \end{equation}
where for the mass of PBHs in the range of $[10^{14},10^{23}]$g results in $\ln\Lambda\in[24,45]$.

We note that the dispersion velocity of matter in the neutron star also is given by the uncertainty principle and exclusion principle for the neutrons (i.e. $pd = \hbar$) where $p$ is the momentum and $d$ is the distance between the neutrons \cite{paddy}. For the ultra-relativistic regime for the neutrons where $E\simeq pc$ and we can set the dispersion velocity of neutrons as $\sigma_n \sim c$. Substituting in definition of $X$ , $$X = 0.22 (\frac{r_s}{1\text{km}})^{1/2}(\frac{r_n}{10\text{km}})^{-1/2}.$$ 
and using equation (\ref{dfric}), we define the time scale for dissipation energy of PBH from the dynamical friction as $t_{df} = v_{pbh}/\dot{v}_{pbh}$ where the numerical value is 
\begin{equation}
    t_{df} = 1.4\times 10^{5}\text{s}(\frac{m_{bh}}{10^{23}\text{g}})^{-1}(\frac{r_n}{10\text{km}})^{3/2}(\frac{M_n}{M_\odot})^{1/2}.
\end{equation}
For the lower and upper bands of PBH mass, the dynamical friction time scale ranges from  $ 10^{5}$ to $10^{14}$ second.

We can compare this time scale with the crossing time scale of PBH across the neutron star by $t_c = r_n/v_{pbh}$ which is 
\begin{equation}
\label{tc}
    t_c \simeq 10^{-4} \text{s} (\frac{M_n}{M_\odot})^{-1/2}(\frac{r_n}{10\text{km}})^{3/2},
\end{equation}
this time scale is very small compared to the dynamical friction time scale and PBHs can not lose a significant amount of their kinetic energy during the crossing of the neutron stars.


Another important parameter in the collision of PBH-neutron star is the 
energy release as a result of the collision. 
 From the dynamical friction, we expect the energy release would be $Q_{df} = -m_{pbh}\dot{v}_{pbh}r_n$. Substituting deceleration from equation (\ref{dfric}), the energy release is:
\begin{equation}
    Q_{df} = - \frac{3}{4}\ln\Lambda \left(erf(X)-\frac{2X}{\sqrt{\pi}}e^{-X}\right)(\frac{r_{s(pbh)}}{r_n})m_{pbh} c^2  
\end{equation}
where using the numerical values the energy released as a result of dynamical friction is 
\begin{equation}
\label{df2}
    Q_{df} = -5.4\times 10^{32} \text{erg} (\frac{m_{pbh}}{10^{23}\text{g}})^2(\frac{r_n}{10 \text{km}})^{-1}.
\end{equation}
We divide this energy by the crossing time scale in equation (\ref{tc}) which results in the power released from the dynamical friction inside the neutron star
\begin{equation}
    P_{df} = 5.4\times 10^{36}\text{erg/s} (\frac{m_{pbh}}{10^{23}\text{g}})^2(\frac{r_n}{10 \text{km}})^{-5/2}(\frac{M_n}{M_\odot})^{1/2}.
\end{equation}

The other source of dissipation is the accretion of the neutron star's material by a PBH during the crossing of the interior of the neutron star. The maximum flux from the accretion is given by the Eddington limit 
\begin{equation}
    L_{edd} = \frac{4\pi G m_{pbh} m_p}{\sigma_{TH}} = 6.5 \times 10^{27}\text{erg/s}(\frac{m_{pbh}}{10^{23}\text{g}}),
\end{equation}
where $\sigma_{TH}$ is the Thompson cross section for photon-electron scattering and $m_p$ is the mass of proton. We note that in both two mechanisms of energy dissipation by the dynamical friction and the accretion, the majority of energy release happens at the interior of the neutron star. This burst of energy can propagate to the surface of the star with the sound speed and the result would be perturbation of the inertial tensor which may cause a kind of glitch or temporary change in the spin of the neutron star. The anomalies in the spin of the pulsar and burst of energy have already been observed \cite{anomaly} and to interpret this observation with a PBH collision, detailed modeling is needed. 

Finally, we investigate the effect of energy dissipation of a PBH after crossing a neutron star. Let us divide $Q_{df}$ from equation (\ref{df2}) (which plays the major role in the dissipation) by the total kinetic energy of a PBH (i.e. $E_k = m_{pbh}v^2/2$), $\epsilon = Q_{df}/E_k$. The result is 
\begin{equation}
 \epsilon \simeq - (\frac{v}{1~\text{km/s}})^{-2} (\frac{m_{pbh}}{10^{23}\text{g}})(\frac{r_n}{10 \text{km}})^{-1}
\end{equation}
In order to have confined PBH around the neutron star, $|\epsilon|\geq 1$ should be satisfied (let us define this velocity as $ v_{th} = v$ with the condition of $|\epsilon|\geq 1$). This means that PBHs with the initial velocity less than $v_{th}$ in the halo can trap around the neutron star, where

\begin{equation}
v_{th} = 1~\text{km/s} (\frac{m_{pbh}}{10^{23}\text{g}})^{1/2}(\frac{r_n}{10 \text{km}})^{-1/2}
\end{equation}

Since the Galactic structures have different dispersion velocities, we expect that the fraction of the PBHs can be captured by the neutron stars depends on the corresponding dispersion velocity of the structures. Here we assume the Maxwell-Boltzmann distribution for velocity of PBHs and the capturing fraction can be evaluated from the integral
\begin{equation}
P(\sigma,m_{pbh}|v<v_{th}) = 1- \sqrt{\frac{2}{\pi}}\int_{v_{th}}^{\infty} \frac{1}{\sigma^3}\exp(-\frac{v^2}{2\sigma^2})v^2 dv, 
\label{int}
\end{equation}
which depends on $\sigma$ as the dispersion velocity of PBHs inside the Galactic structure and the mass of PBHs. 

\begin{figure}[pb]
\centerline{\psfig{file=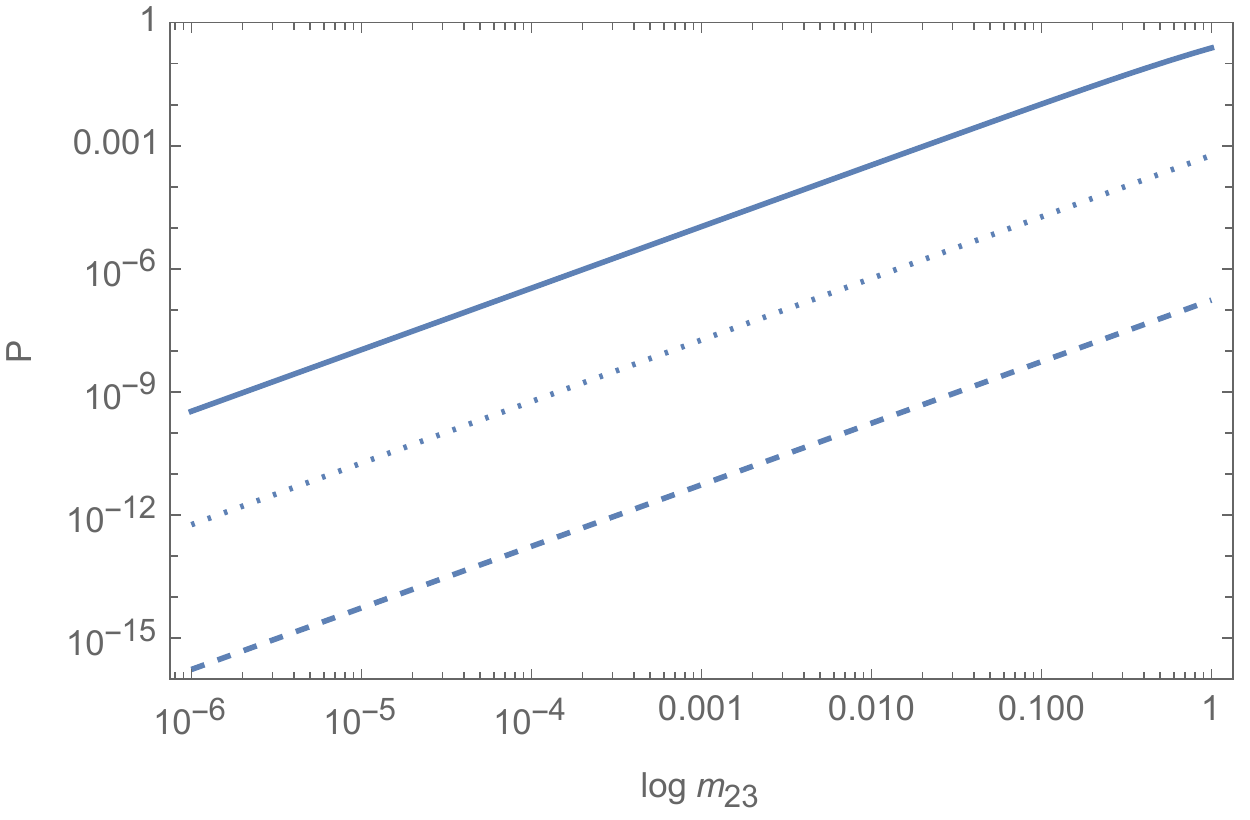,width=8.7cm}}
\vspace*{8pt}
 \caption{ This graph shows the likelihood of a PBH (Primordial Black Hole) being trapped in a neutron star. The graph plots the probability against the normalized mass of 
  $m_{23} = (m/10^{23}\text{g})$ and includes three models: a halo model (dashed line), a dark disk model 
with $\sigma = 80 \text{km/s}$ (dotted line), and a dark disk model with $\sigma = 40 \text{km/s}$ (solid line).
  }
  \label{fig1}
\end{figure}

\begin{figure}[pb]
\centerline{\psfig{file=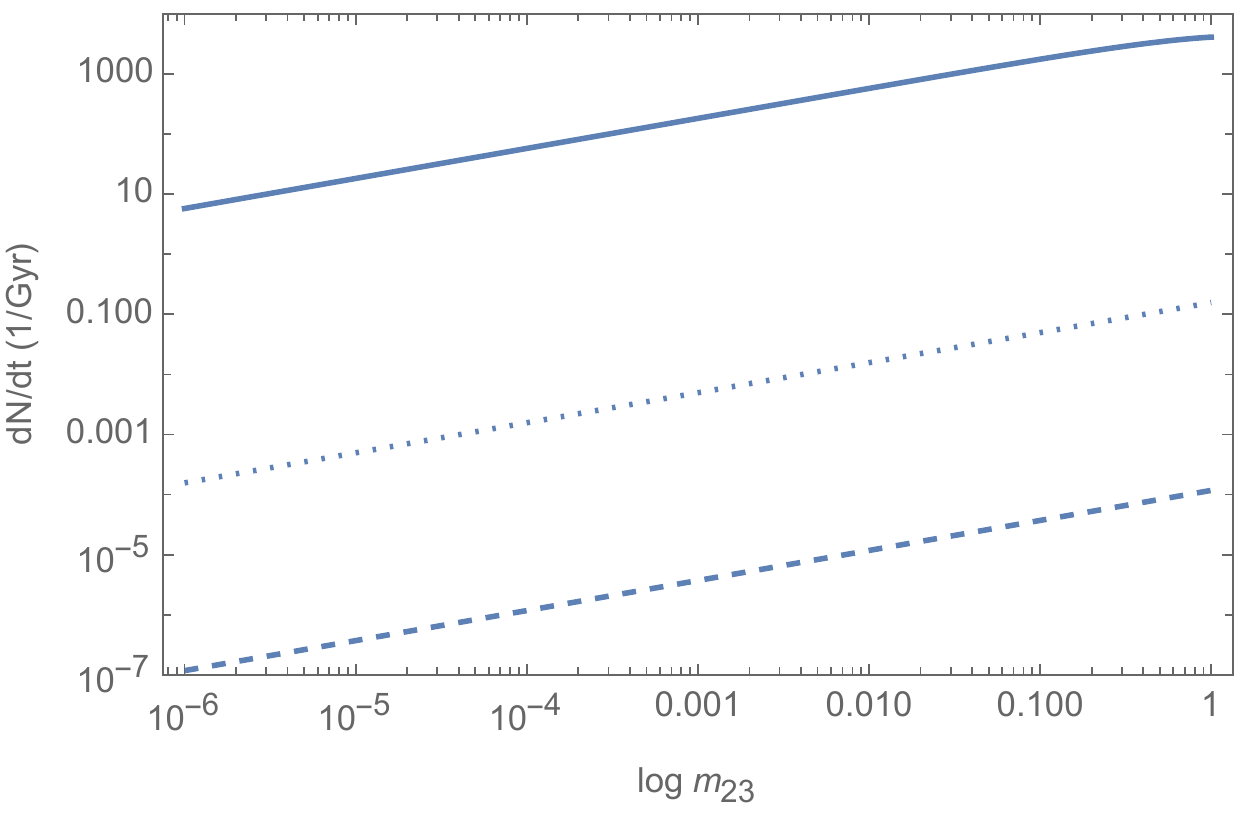,width=8.7cm}}
\vspace*{8pt}
  \caption{The rate of PBHs of Milky Way captured in a neutron star as a function of normalized mass of $m_{23}=(m/10^{23} \text{g})$ for three different models of the halo (dashed line), dark disk with $\sigma = 80 \text{km/s}$ (dotted line) and dark disk with $\sigma = 40 \text{km/s}$ (solid line) in the Milky Way. The estimated total number of neutron stars in the Milky Way is $10^8$ stars  \cite{Reed_2021}.}
  \label{fig2}
\end{figure}

The dispersion velocity of particles inside the Galactic halo is $\sigma_\simeq 200 \text{km/s}$. Also there is thin to thick dark disk models with the dispersion velocities in the range of $\sigma = 40-80 \text{km/s}$ \cite{purcell}. Figure (\ref{fig1}) represents the result of the integration of equation (\ref{int}) for various dark components of the Galaxy. 
Now, we multiply this probability function by the rate of PBH impact with a neutron star from equation (\ref{rate4}) to obtain the rate of a PBH capture for different Milky Way dark structures. The estimation for the total number of neutron stars in the Milky Way \cite{Reed_2021}  is $N_n \simeq 10^8$. By multiplying the numerical values for the probability, rate and number of stars (i.e. $ \frac{dN}{dt}\times P(v<v_{th})\times N_n$), we obtain the PBH capturing rate by the neutron stars for different dark components of the Milky Way as shown in Figure (\ref{fig2}).

The capturing rate, measured in  Gyr$^{-1}$, makes it unlikely for us to witness these events in the Milky Way (except for the low mass PBHs, Figure \ref{fig2}) . However, on a cosmological scale, along our past light-cone there is potential for these events to be detected by observing large number of galaxies at higher redshifts.

When a PBH is captured by a neutron star, it can either merge with the star resulting in energetic outbursts, or be put in an orbital motion around the star. The outcome depends on the impact parameter of the collision between the PBH and the neutron star.


\section{Physical consequence of PBH collision with the Astrophysical black holes}
\label{sec4}

In this section, we study the gravitational collision between PBHs and astrophysical black holes, which can be detected through the propagation of gravitational waves. We note that the PBHs have been made in the early Universe, while the astrophysical black holes are the final stage of the heavy stars evolve to a black hole.  Previous research \cite{GW1,GW2,GW3,GW4}, has extensively studied the signals produced by the interaction of PBH-PBH and PBH-Astrophysical black holes.

We provide a basic estimation of the energy released from this collision and study the observability of this event by estimating the energy release resulting from the merger of two spin-less black holes with masses $M$ and $m_{pbh}$
under head-on collision  \cite{hawking2} and forming a black hole with mass $M_t$ . Using the definition of the ADM-mass \cite{adm}, we can express $(m_{pbh} + M)c^2 = E_{gw} + M_tc^2$ where $E_{gw}$ is the energy released by the gravitational waves. On the other hand, the Entropy of a black hole is proportional to the area of a black hole (i.e. $S\propto A \propto M^2$). Thus, according to the second law of thermodynamics $S(M_t) >S(m_{bh}) + S(M_n)$, or in another word, $M_t^2>m_{pbh}^2 + M^2$. Combining the conservation of energy-momentum with the second law of thermodynamics constrains the energy of gravitational wave from this collision as 
\begin{equation}
\label{egw0}
    E_{gw}< (m_{pbh} + M - \sqrt{m_{pbh}^2 + M^2})c^2,
\end{equation}
where in our case $m_{pbh}\ll M$ and equation (\ref{egw0}) simplifies to  
\begin{equation}
    E_{gw}< m_{pbh}c^2 - m_{pbh}c^2 \frac{m_{pbh}}{M}.  
\end{equation}

Gravitational wave emission from the collision of an object with a black hole has been analyzed in detail \cite{ruffini}. The energy of gravitational wave emission is given by
\begin{equation}
\label{egw}
    E_{gw} 
    = 5\times 10^{31}\text{erg}(\frac{m_{pbh}}{10^{23}\text{g}})^2(\frac{M}{10M_\odot})^{-1},
\end{equation}
where the total spectrum of gravitational wave peaks at the angular frequency of 
\begin{equation}
\label{omega}
    \omega = 0.32 \frac{c^3}{GM} = 0.6\times 10^4 \text{Hz} (\frac{M}{10M_\odot})^{-1}.
\end{equation}
Multiplying equation (\ref{egw}) to (\ref{omega}) results in an estimation for the power of the gravitational wave,  
\begin{equation}
\label{power}
P_{gw} \simeq 0.3 \times 10^{36}\text{erg/s} (\frac{M}{10M_\odot})^{-2} (\frac{m_{pbh}}{10^{23}\text{g}})^2.
\end{equation}

Now, we estimate the amplitude of gravitational waves from this system at a given distance from the Earth. The energy-momentum of gravitational waves averaged over the several wavelengths \cite{mag} is given by 
\begin{equation}
\label{t}
    T^{\mu\nu} = \frac{c^4}{32\pi G}<\partial^\mu h^{\alpha\beta}\partial^\nu h_{\alpha\beta}>,
\end{equation}
where energy-momentum satisfies the conservation laws in Minkowski background (i.e.  $T^{\mu\nu}{}_{,\nu} = 0$).     Integrating over the spatial volume for $\mu = 0$ component is $\int T^{00}{}_{,0} d^3x = - \int T^{0i} ds_i$. Here the left-hand side of this equation is the power of gravitational wave as in equation (\ref{power}) and the right-hand side can be substituted from equation (\ref{t}). For a plane wave at far distances from the source $h^{\alpha\beta} \propto \exp(k\cdot r - \omega t)$, we obtain the amplitude of the gravitational wave in terms of power and distance of the source from the observer as
\begin{equation}
\label{h1}
    |h| \simeq \frac{3 GM}{r c^4}(\frac{8 G P}{c})^{1/2}.
\end{equation}
We substitute the gravitational wave power from equation (\ref{power}). The final result for the amplitude of $h$ is 
\begin{equation}
\label{h3}
|h| \simeq 4\times 10^{-27}(\frac{m_{pbh}}{10^{23}\text{g}})(\frac{r}{1\text{kpc}})^{-1}.
\end{equation}
The interesting feature of this equation is that the amplitude of gravitational wave emission from PBH-astrophysical black hole collision is independent of the mass of the astrophysical black hole. The physical reason is that the gravitational wave power is proportional to $1/M^2$ from equation (\ref{power});  substituting in (\ref{h1}) results in a gravitational wave amplitude independent from the mass of the astrophysical black hole. We note that in all the calculation we assumed that $M\gg m_{pbh}$.   

From equation (\ref{h3}), the amplitude of the gravitational wave at the observer's position is three orders of magnitude smaller than the sensitivity of the Advanced LIGO detector \cite{aligo}. Therefore, detecting PBH-astrophysical black hole collisions remains a target for future gravitational wave detectors \cite{luca}.


\section{Conclusion}
\label{sec5}

Summarizing this work, we investigated the probability and the physical consequences of collision of the Primordial Black Holes (PBHs) as the candidate for dark matter with compact objects. The compact object in our study could be either a neutron star or an astrophysical black hole. Using a Schwarzschild metric for the compact object as neutron star or an astrophysical black hole, we obtained relativistic trajectories of PBHs around these objects.  The collision rate of PBHs with compact objects depends on the mass function of PBHs and the mass of compact objects. We estimate the amount of dissipation during the collision of PBHs with the neutron stars. The time scale of dissipation is much larger than the PBH crossing time scale of the neutron stars. However, having a Maxwell-Boltzmann distribution for the velocity of PBHs, a small fraction of them can be captured in the neutron stars, either by merging the neutron star or staying in an orbit around the neutron star. The total rate of this capturing is very low and it is unlike to be detected in the Milky Way during our lifetime. However, signals for this event on the cosmological scale are expected. Also PBHs orbiting around the neutron stars and black holes might be a target for the gravitational microlensing  
observations \cite{rahvar_mic} through binary microlensing channel. However, we note that since the rate of binary formation according to Figure (\ref{fig2}) is low, we would expect that for a Milky Way type galaxy, there will be almost $10$ PBH orbiting around the compact objects. However if the primordial black holes theories let existence of smaller mass PBHs, then the number of neutron star-PBH binaries would be higher and one could reveal the existence of PBHs though the microlensing observations.


We also investigated the consequence of the collision of PBH with the astrophysical black holes. We showed that the amplitude of the gravitational wave signal for a PBH-astrophysical black hole collision located at kilo parsec distance is 
three orders of magnitude smaller than the sensitivity of the present-day detectors. Future detectors may observe these waves as a signature for the existence of PBHs in the halo of our Galaxy.

\nocite{*}



\section*{Acknowledgements}
We would like to thank anonymous referee for his/her useful comments improving this work. 

 



\bibliographystyle{ijmpd}
\bibliography{example} 




\appendix




\label{lastpage}
\end{document}